\documentclass[%
 aip,
 rsi,
 amsmath,amssymb,
 reprint,%
]{revtex4-1}

\usepackage{graphicx}
\usepackage{dcolumn}
\usepackage{bm}

\usepackage[utf8]{inputenc}
\usepackage[T1]{fontenc}
\usepackage{mathptmx}
\usepackage{hyperref}
\begin{document}

\preprint{AIP/123-QED}

\title[Efficient Water-Cooled Bitter-Type Electromagnet for Zeeman Slowing in Cold-Atom Experiments]{Efficient Water-Cooled Bitter-Type Electromagnet for Zeeman Slowing in Cold-Atom Experiments}

\author{Rishav Koirala}
\affiliation{Centre for Quantum Technologies, National University of Singapore, 117543, Singapore}
\author{Ben A. Olsen}%
 \homepage{https://olsenlab.science}
 \email{bolsen@lclark.edu}
\affiliation{ 
Lewis \& Clark College, Department of Physics, Portland OR 97219, USA
}%

\date{\today}

\begin{abstract}
We describe the design, construction, and characterization of a Bitter-type electromagnet that produces a spatially-dependent magnetic field used for Zeeman slowing in cold-atom experiments. The coil consists of stacked copper arcs separated by PTFE spacers of varying thicknesses, generating a near-optimal field profile using a single power supply. With an electrical resistance of $26.5(3)$~m$\Omega$ and self-inductance of $19.1(1)$~$\mu$H, our design achieves a fast electrical switching time of $\tau \approx 100$~$\mu$s in a compact, 30-cm-long package. Water circulating helically through holes in the copper and channels in the spacers ensures efficient thermal management, limiting the temperature rise to $\sim 5^\circ$~C over $36$~s of continuous operation at $200$~A.

\end{abstract}

\maketitle

\section{\label{sec:intro}Introduction}

The Zeeman Slower (ZS)\cite{phillipsLaserDecelerationAtomic1982} is a key component of many laser-cooling experiments, where a fast beam of atoms, often emitted from a thermal source, is slowed enough to be captured in a Magneto-Optical Trap (MOT)\cite{raabTrappingNeutralSodium1987}. 
A key ingredient of the ZS is a magnet, conventionally an electromagnet shaped like a hollow cylinder surrounding the atom path, that produces an inhomogeneous magnetic field profile. 
The position-dependent Zeeman shift of the atomic transition from this field profile is engineered to cancel the Doppler shift associated with the atomic motion, allowing the slowed atoms to continuously scatter photons from a single-frequency laser beam along the entire length of the ZS\cite{firminoProcessStoppingAtoms1990}. 
The resulting acceleration produces a roughly 100-fold reduction in the velocity of the slowed atoms.

To produce the appropriate inhomogeneous magnetic field profile, several designs have emerged, in three popular field profile/laser polarization combinations: decreasing-field/$\sigma^+$-polarized, increasing-field/$\sigma^-$-polarized, and field-crossing (or ``spin-flip'').
The most prevalent magnet for all three field types is a wire-wound electromagnet coil, consisting of several segments of insulated copper wire wrapped concentrically into solenoid segments with varying numbers of turns or layers\cite{phillipsLaserDecelerationAtomic1982,firminoProcessStoppingAtoms1990, barrettSlowingAtomsPolarized1991, martiTwoelementZeemanSlower2010,dedmanOptimumDesignConstruction2004,ohayonInvestigationDifferentMagnetic2015a}.
They produce a longitudinal magnetic field, and due to the large number of turns of wire, require modest currents (though many designs require several power supplies to drive unequal currents in various segments of the ZS coil).
These wire-wound coils are typically water-cooled with layers of hollow tube, and the entire assembly is integrated with an ultra-high vacuum (UHV) chamber.

In one simplification to this wire-wound design, the cooling lines were combined with the current-carrying wires, and a single layer of hollow-core wire was wound around a cylinder with variable pitch angle to produce the field profile\cite{bellSlowAtomSource2010}.
This design requires only a single power supply, and has much lower electrical resistance and self-inductance compared to traditional designs, with a field-switching time of around 0.3~ms.

Another alternative method to produce the required field profile is to use an array of permanent magnets.
Some designs employ a Halbach array of magnets with varying transverse spacing\cite{cheineyZeemanSlowerDesign2011, aliDetailedStudyTransverse2017, wodeyRobustHighfluxSource2021a}, while others use a series of cylindrical permanent magnets with their distance from the atom beam individually adjusted with screws\cite{hillZeemanSlowersStrontium2014, yuZeemanSlowingGroupIII2022}, layers of self-assembled spherical permanent magnets\cite{lebedevSelfassembledZeemanSlower2014}, rectangular permanent magnets held in place with a 3D-printed polymer form\cite{parsagianDesigningBuildingPermanent2015} or on a CNC-milled metal form \cite{liIntegratedHighfluxCold2023}, or a series of ring-shaped permanent magnets with varying inner diameter and thickness\cite{wangLongitudinalZeemanSlower2015} (which produces a longitudinal field).
One hybrid slower design combines an electromagnet with a permanent magnet array to increase the effective length of the field profile\cite{garwoodHybridZeemanSlower2022}.

Each of these designs involve certain tradeoffs: permanent-magnet designs require no power supplies, and generate no heat, while electromagnet-based designs can be turned off to eliminate fringe fields at the MOT location.
Most permanent-magnet designs produce a magnetic field transverse to the atomic beam, which necessitates an additional optical pumping preparation stage \cite{aliDetailedStudyTransverse2017}.
All wire-wound designs are captured by the UHV chamber, making them difficult or impossible to reconfigure or repair.
Designs with several segments or layers offer increased adjustability of the field profile using multiple independent power supplies\cite{ohayonInvestigationDifferentMagnetic2015a}, at the cost of increased laboratory equipment and more complicated construction.

In designing our experiential apparatus, we prioritized the ability to turn off the field quickly while minimizing power dissipation.
Our overall apparatus precluded a variable-pitch design like\cite{bellSlowAtomSource2010}, due to a combination of limited manufacturing capabilities and a shorter planned ZS length. 
To overcome some of the limitations of wire-wound electromagnets, while retaining field-switching speed, we chose an alternative electromagnet coil design.
We opted for a Bitter-type electromagnet coil made of stacked layers of copper \cite{bitterDesignPowerfulElectromagnets1936, bitterDesignPowerfulElectromagnets1939}.
This choice was inspired by controllable electromagnet designs in other cold-atom experiments, where Bitter-type coils were used to produce homogeneous bias fields \cite{sabulskyEfficientContinuousdutyBittertype2013, longAllopticalProductionLi62018}, bias fields with gradients \cite{luanModifiedBittertypeElectromagnet2014}, or a combination of bias, gradient and curvature fields \cite{siegelBittertypeElectromagnetComplex2021a}.
These Bitter-type coils have fewer turns than wire-wound designs, with larger cross-section, leading to lower self-inductance and resistance, and the coils can fit in constrained geometries.

In this article, we describe the design, construction, and characterization of a Bitter-type electromagnet to produce the inhomogeneous field profile for Zeeman slowing of a beam of neutral lithium atoms.
Modifying the traditional Bitter design by incorporating layers and inter-layer spacers, both with with varying thickness, we engineer a near-ideal ZS field profile.
We characterize the coil's electromagnetic properties, including its self-inductance, resistance, its magnetic field profile, as well as its field switching time.
We also investigate the coil's thermal performance, including thermalization times and temperature profile. Finally, we discuss the limitations of our coil design and offer some suggestions for further improvements.



\section{\label{sec:zs}Zeeman Slower Coil Design}

\subsection{\label{ssec:profile}Magnetic Field Profile}

For a decreasing-field ZS of axial length $l$, atoms moving toward $+z$, and a counter-propagating laser with constant wavelength $\lambda$ and frequency detuning $\delta$ from the atomic transition, the ideal magnetic field profile leading to constant deceleration of the atoms to rest has the form
\begin{equation}\label{eqn:profile}
    B(z) = B_\textrm{bias} + B_0\sqrt{1 - \frac{z}{l}},
\end{equation}
where $B_\textrm{bias} = h|\delta|/\mu_B$ is a constant offset field, $B_0 = h v_p/\lambda \mu_B$ is the peak strength of the field measured from this offset, and $z$ is the axial position along the ZS. The physical constants $\mu_B$ and $h$ are the Bohr magneton and the Planck constant, respectively.
The lithium atoms effusing out of an oven at $T=670$~K\cite{mitraExploringAttractivelyInteracting2018,huletMethodsPreparingQuantum2020}  have a most-probable velocity of $v_p=\sqrt{3k_BT/m}\approx1670$~m/s from the Maxwell-Boltzmann distribution, where $k_B$ is the Boltzmann constant and $m$ is the mass of a $^6$Li atom. 
The laser light at $\lambda = 671$~nm (targeting the D2 line of $^6$Li) has a detuning $\delta$ that is adjusted in the experiment to maximize the final atom number in the MOT. 
For a nominal $\delta = 2\pi \times -100$~MHz, we obtain $B_0 = 0.18$~T and $B_\text{bias} = 0.04$~T. 
The atoms scatter photons from the laser beam at a rate determined by the natural linewidth $\Gamma=2\pi \times 5.9$~MHz and experience a maximum deceleration of $a_D=h \Gamma / 2m \lambda =1.8\times 10^6$ m/s$^2$ before coming to a stop over a distance of $l=v_p^2/2a_D=0.76$~m.
If all atoms with initial velocity less than $v_p$ are slowed, this would lead to $44\%$ of the atoms in the beam slowed.

In practice, however, several technical considerations can modify the distance, leading to a wide range of ZS lengths. 
For example, atoms usually decelerate with $a < a_D$ due to limited laser power, which can increase $l$ by up to a factor of two. 
On the other hand, they often do not need to be slowed down completely, but only until their speed $v < v_c$, the capture velocity of the MOT, reducing the required slowing distance.
Other lithium experiments with wire-wound coils have $l$ of $\approx 0.35$~m\cite{huletMethodsPreparingQuantum2020, garwoodHybridZeemanSlower2022, mitraExploringAttractivelyInteracting2018}, $\approx 0.7$~m\cite{longAllopticalProductionLi62018}, and up to $\approx 1$~m\cite{martiTwoelementZeemanSlower2010, wangColdControlledLithium2024} (usually for slowers for two atomic species). 
While a shorter length ZS yields smaller deceleration, leading to a smaller slowed fraction of the beam, this loss is compensated for by the increased solid angle subtended by the output aperture of the shorter ZS\cite{huletMethodsPreparingQuantum2020}.
Several recent experiments have combined the magnetic field produced by the MOT and ZS coils to produce the optimal profile\cite{garwoodHybridZeemanSlower2022,mitraExploringAttractivelyInteracting2018}.
This configuration reduces the influence of the transverse motion of the slowed beam; by the time the atoms have slowed to the capture velocity of the MOT, they are already within the MOT capture region (within $\approx 1$~cm of where $B=0$).

\subsection{\label{ssec:coil}Bitter-type electromagnet coils}

\begin{figure}[ht]
    \centering
    \includegraphics[width=\linewidth]{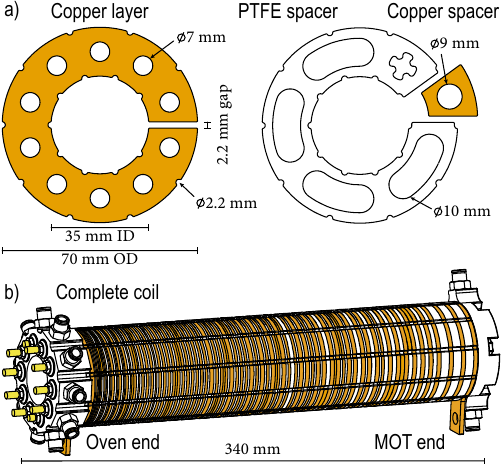}
    \caption{Geometry of the coil parts. a) Each conducting layer is an annular piece of OFHC copper of constant thickness, with a single gap, 10 holes for axial cooling water flow, and small alignment notches. Between the layers, a PTFE spacer and copper spacer of equal thickness fit together. The PTFE spacer has 4 channels for azimuthal cooling water flow, and one clover-shaped hole to admit axial cooling water flow, while preventing the threaded rod from contacting the neighboring copper layers. For thin copper spacers, a rubber gasket fits inside the hole and seals the neighboring layers from cooling water leaks. For thicker layers, a thin channel is cut out around a 7~mm hole on both sides of the spacer, so two gaskets form a seal with the neighboring layers. b) The overall coil has 71 copper layers, separated by PTFE spacers, and is held together with axial tension provided by 10 threaded rods. The PEEK endcaps have connections for cooling water---10 at the oven end of the coil, and 2 at the MOT end.}
    \label{fig:geometry}
\end{figure}

\begin{figure*}[ht]
    \centering
    \includegraphics[width=\textwidth]{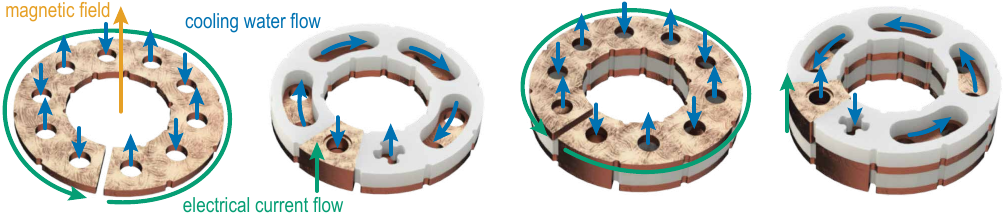}
    \caption{Schematic of coil layer construction. In one copper layer, the azimuthal electric current, indicated with a green arrow, flows counter-clockwise, and cooling water flows axially in the direction indicated by blue arrows. In the spacer between copper layers, a copper spacer carries electrical current axially, while channels in the PTFE insulator guide the water to flow azimuthally. The strength of the axial magnetic field in the coil center, shown in orange, is controlled by varying the layer and spacer thicknesses.}
    \label{fig:flow}
\end{figure*}

For electromagnets that dissipate a fixed amount of power, Bitter showed that a coil made from stacked, split annular layers of conductor was close to optimum for generating high fields, while still being practical to construct\cite{bitterDesignPowerfulElectromagnets1936,bitterDesignPowerfulElectromagnets1939}.
This Bitter-type coil geometry is especially useful in space-constrained applications, such as re-entrant viewports on a UHV chamber\cite{sabulskyEfficientContinuousdutyBittertype2013, longAllopticalProductionLi62018}.
Although most Bitter-type coils are designed to produce a uniform field over the largest volume possible, we aimed to keep the benefits of the stacked layer design while creating the field profile of Eq.~\ref{eqn:profile}, with maximum value $B_0\approx 70$~mT, and characteristic length $\ell\approx 0.3$~m, similar to\cite{mitraExploringAttractivelyInteracting2018}.

Because we aimed to use the combined field of the ZS and MOT coils to produce this profile, the ZS coil would produce a significant field at the center of the MOT.
During a typical cooling sequence for lithium,  MOT loading is followed by Gray Molasses (GM) cooling, which requires zero magnetic field over the entire $\approx 1$~cm atom cloud, and occurs $0.2$--$0.5$~ms after the MOT coils are turned off\cite{satterComparisonEfficientImplementation2018}.
This sequence requires the ZS coil to also turn off quickly, requiring the coil design to have the lowest possible self-inductance $L$.
Bitter-type coils have fewer current loops $N$ than similar sized wire-wound coils, and since the self-inductance for a solenoid scales as $L\propto N^2$, a Bitter-type coil should turn off more quickly.

To produce the same amplitude field with fewer turns, the Bitter-type coil requires a  higher current $I$, as $B\propto N I$.
Fortunately, commercially available DC power supplies can provide currents up to 400~A at moderate cost.
As a consequence, minimizing power dissipation $P=I^2R$ requires a coil with the lowest possible resistance $R$.
As we will see in Sec.~\ref{sec:thermal}, this resistive heating can be balanced by water cooling.

To maximize the field strength, the coil's inner radius is as close as possible to the outer radius of a tube on the UHV chamber with a 1.33" CF flange.
The entire UHV chamber must be baked to reach background gas pressures low enough for laser-cooling experiments, so the entire ZS coil assembly must be able to withstand bake-out temperatures up to $100^\circ$~C.






\subsubsection{Coil Design}

To design the ZS coil, we began by modeling the basic geometry of the conductor in {\scshape radia}\cite{chubarThreedimensionalMagnetostaticsComputer1998}, a 3D magnetostatics software package for {\scshape mathematica} or {\scshape python} that can simulate the magnetic field produced by distributed current densities of various geometries. 
The two building blocks for the conducting parts of the coil are layers and spacers.
Each layer has a rectangular cross-section, and swepdf out a near-circular path, producing a `C'-shaped disc, as shown in Fig.~\ref{fig:geometry}.
Each spacer has constant thickness and is bounded by an inner and an outer radius, as well as two radial arcs separated by $2\pi/10=36^\circ$, forming a near-trapezoidal prism.
For each copper spacer, we also designed an insulating PTFE spacer of the same thickness that occupied the other $324^\circ$ arc between layers.

\begin{figure}[ht]
    \centering
    \includegraphics[width=\linewidth]{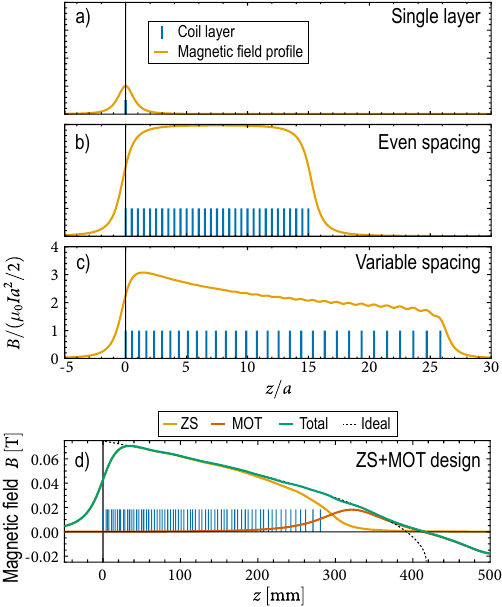}
    \caption{Iterative field profile design. (a) a single layer produces a field profile similar to a loop of current, $B(z)\propto a^2/(z^2+a^2)^{3/2}$, where $a$ is the loop radius. The location of the layer is shown with a vertical blue line segment. (b) a collection of many identical layers and spacers produces a nearly constant field profile, similar to a solenoid. (c) increasing the spacing of some of the layers and spacers decreases the local field strength. (d) after several rounds of manually changing the layer spacings among the set of stock thicknesses, we designed a ZS coil to produce a nearly-ideal field profile (green curve) when combined with the field produced by a pair of MOT coils (red curve).}
    \label{fig:coil}
\end{figure}

In our {\scshape radia} simulation, we used a slightly simplified geometry: for the coils, a constant current density flowed azimuthally through an angle of $358^\circ$, and for the spacers, a constant current density flowed axially through a rectangle with the same cross-sectional area.
Each layer is rotated from the previous layer by $36^\circ$, with a spacer connecting opposite azimuthal ends of the layers, so the current in each layer flows clockwise (See Fig.~\ref{fig:flow}), producing a net magnetic field that is primarily axial.

For fixed inner and outer radii, the magnetic field profile $B(z)$ along the axis of revolution (the path of the atomic beam) is determined by the layer and spacer thicknesses.
To keep costs low, we considered only thicknesses of oxygen-free high-conductivity (OFHC) copper (alloy 101) available from commercial suppliers: 0.025", 0.032", 0.04", 0.05", 0.063", 0.08", 0.093", 1/8", 3/16", 1/4", 3/8", 1/2".
To reduce the complexity of the coil, we further restricted to 3 layer thicknesses (0.04", 0.08", and 1/8") and 4 spacer thicknesses (0.04", 0.08", 1/8", and 1/4").

To design the field profile in our simulation, we began with a large number of coils and spacers of the minimum thickness, which produce a nearly constant magnetic field, similar to a solenoid (See Fig.~\ref{fig:coil}).
We then iteratively increased the thicknesses of some of the layers and spacers, beginning with the layers farthest from the beam source (at higher $z$).
Increasing a layer's thickness reduces the current density, as well as increasing the spacing between neighboring layers, both of which lower the peak value of the magnetic field in the vicinity of the layer.
Increasing a spacer's thickness does not change its current density, but increases the distance between neighboring layers, reducing the local field strength.
By comparing the simulated field profile to an ideal profile with the same maximum field strength, we could adjust the number and spacing of the layers until the profiles matched very well.
A table of the final layer and spacer thicknesses can be found in the design files, along with a {\scshape radia} simulation, bill of materials, and raw data at \href{https://github.com/olsenlab-science/Bitter-ZS}{https://github.com/olsenlab-science/Bitter-ZS}.

To hold the coils together, we employed a system of 10 threaded rods with nuts and washers on both ends of the coils to provide axial compression forces (see Fig.~\ref{fig:geometry}b), inspired by previous bitter-type coil designs\cite{sabulskyEfficientContinuousdutyBittertype2013}.
To provide space for these threaded rods, we removed 10 circular holes from each layer and 1 hole from each copper spacer.
To prevent excess heating, we also used a water-cooling system for the coil, where the cooling water flows axially through the layers and copper spacers, then azimuthally through channels in the PTFE spacers (see Fig.~\ref{fig:flow}).
This flow pattern balances mechanical stability against compression with increased water-copper contact area.
On the endcaps of the coil (made from rigid, electrically insulating PEEK), the input and output water cooling lines protrude radially outward through adapters to flexible tubes.
In our experiment, the MOT coils and optical access through the chamber viewports constrain that end of the coil to only two tubes, while the other end has 10 tubes.
To minimize the distance from the coil to the center of the MOT chamber, we added recessed holes for the nuts and washers on that end of the coil so the endcap would be flush against the UHV chamber (the right end of Fig.~\ref{fig:geometry}c).

\subsubsection{Coil Construction}

Since each layer and spacer has constant axial cross-section, we cut them from sheet stock, using a 3-axis CNC mill for the copper parts and a laser cutter for the PTFE parts.
The endcaps have more complicated geometry, so we had them cut at a small commercial prototype CNC shop.
To assemble the coil, we built a jig to hold the threaded rods vertically, and hand-placed each part beginning with the thicker layers near the MOT end of the coil.
Since the nuts at the MOT end/bottom of the coil were recessed in the endcap, we could not apply torque to them for tightening.
Instead, we applied torque to the threaded rod using an extra pair of counter-tightened nuts above the upper nuts.
Since the PTFE spacers deflect slightly, we needed to re-tighten the nuts over several hours.

Once the coil was tightened, we attached the cooiling water lines, with about a 340~kPa (50~psi) pressure difference between input and output.
In the first version of the coil, we found that water would leak out between the interfaces between layers and copper spacers, even though the surfaces were polished smooth (the PTFE-copper interfaces were generally water-tight). 
We modified the design by enlarging the holes in each thinner spacer to accommodate a rubber o-ring of nearly the same thickness.
For the thicker spacers, an o-ring would deform too much, so we instead used the CNC mill to machine out a thin notch for two thin o-rings (one each on the top and bottom of the spacer).
The addition of these o-rings substantially increased the complexity of assembling the coil, and the newer design still had a small number of slow water leaks.
We coated both the inner and outer radial surfaces with a layer of high-temperature silicone sealant, which stopped the water leaks.
For future iterations of this design, we suggest an adhesive between the o-rings and the copper spacers to ease the construction process.

\section{\label{sec:em}Electromagnetic properties} 


\subsection{\label{ssec:simulations}Magnetic Field Profile}

\begin{figure}
    \centering
    \includegraphics[width=\linewidth]{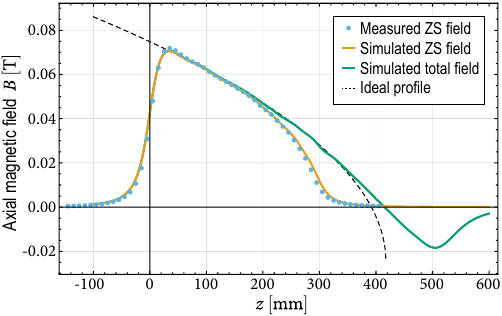}
    \caption{Magnetic field profile of the Bitter ZS coil at $I = 200$~A. The measured field strength along the central axis of the coil is shown in blue dots and agrees well with the predicted field of the ZS based on our simulations (green curve). The minute deviations are likely due to compressive forces slightly decreasing the length of the PTFE spacers. The total magnetic field (yellow curve) includes contributions from the ZS coil, as well as other electromagnet coils on the experimental chamber. It matches the ideal ZS field profile (dotted line) well between $z\approx 40$~mm and $z=380$~mm, roughly 10~mm from the capture volume of the MOT.}
    \label{fig:bfield-profile}
\end{figure}

Using an axial magnetic field probe\footnote{Lakeshore Cryotronics HMMA-2504-VR} fixed in the center of a plastic tube and inserted into the ZS coil, we measured the magnetic field profile $B(z)$ inside and near the coil with current $I=200$~A.
As seen in Fig.~\ref{fig:bfield-profile}, the magnetic field profile agrees well with our simulated predictions.
We also measured one of the radial components of the magnetic field using a transverse-field probe\footnote{Lakeshore Cryotronics HMMA-2504-VR}, and found it vanished to within our measurement uncertainty along the length of the ZS.


\subsection{\label{ssec:impedance}Coil Impedance}

\begin{figure}
    \centering
    \includegraphics[width=\linewidth]{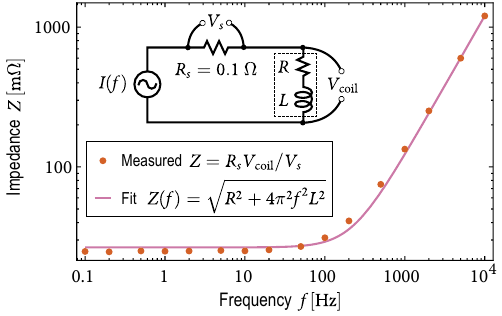}
    \caption{
    Electrical impedance $Z$ of the ZS coil as a function of AC current frequency $f$, measured using the circuit in the inset. A controllable current source drove a sense resistor $R_s$ in series with the ZS coil (modeled as a lumped $RL$ circuit with resistance $R$ and self-inductance $L$).
    The measured impedance $Z(f)= R_s V_\text{coil}/V_s$ (red dots) was fit with the $RL$ series impedance $Z(f)=\sqrt{R^2+4\pi^2f^2L^2}$ (pink curve) to obtain $R = 26.5(3)$~m$\Omega$ and $L=19.1(1)~\mu$H.}
    \label{fig:ZSRL}
\end{figure}

We measured the electrical impedance $Z(f)=V(f)/I(f)$ as a function of frequency $f$ by applying an AC current to the coil, measuring the current $I(f)=V_s/R_s$ using a $R_s=0.1~\Omega$ sense resistor in series, and measuring the voltage $V_\text{coil}(f)$ across the ZS coil, as seen in Fig.~\ref{fig:ZSRL}.
Using a lumped $RL$ model for the coil, the impedance is related to the resistance $R$ and self-inductance $L$ by
\begin{equation}
    \label{eqn:impedance}
    Z(f)=\sqrt{R^2+4\pi^2f^2L^2}.
\end{equation}
We varied the AC frequency over 5 orders of magnitude, as shown in Fig.~\ref{fig:ZSRL}, and fit the measured $Z(f)$ using this model to find $R=26.5(3)$~m$\Omega$ and $L=19.1(1)~\mu$H. 
A similar length wire-wound ZS coil\cite{mitraExploringAttractivelyInteracting2018} had an estimated $R=170$~m$\Omega$. 




\subsection{Field Switching}

\begin{figure}
    \includegraphics[width=\linewidth]{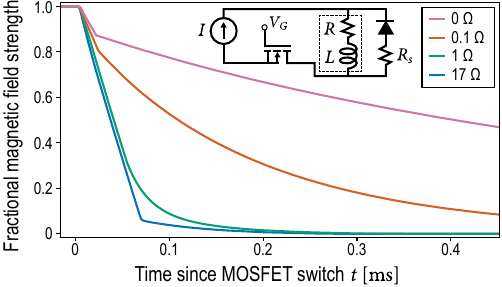}
    \caption{
    Magnetic field decay after the current is rapidly switched off using a MOSFET. 
    Driving the coil with the circuit depicted in the inset, we measured the magnetic field using a multi-loop wire placed in the bore of the coil and integrating the resulting voltage.
    We see an initial fast linear decay, followed by a slower decay, with characteristic time roughly given by $L/(R+R_s)$ for a few values of $R_s$.
    The initial linear decay time is limited by some characteristic $V_\text{max}/L$.}
    \label{fig:ZSimpulse}
\end{figure}

We measured the response of the coil to a rapidly switched current using the circuit depicted in Fig.~\ref{fig:ZSimpulse}, where a MOSFET \footnote{IXYS model IXFN240N25X3} was used as a fast switch.
When the gate voltage $V_G$ is high, the MOSFET conducts, and a DC current $I=200$~A passes through the ZS coil to produce a magnetic field in the bore of the coil.
Once $V_G$ suddenly goes low at $t=0$, current flows through the ZS coil, a shunt resistor, and a diode, similar to \cite{dedmanFastSwitchingMagnetic2001}.
We measured the magnetic field strength in the bore of the ZS coil using a small loop of wire with many turns which has $V\propto d B/dt$, which we integrated to calculate the field decay, shown in Fig.~\ref{fig:ZSimpulse}.
For different values of $R_s$, we see an initial linear decay of the field, followed by a slower decay with characteristic time given roughly by $L/(R+R_s)$. 
For $R_s=17~\Omega$, the field decayed nearly to zero in about $75~\mu$s.
In all cases, the slow decay was not exponential, likely due eddy currents in the coil or nonlinear diode $IV$ characteristics.

Some other cold-atom experiments use more complicated switching circuits, including transient voltage suppressors\cite{Aubin2005a}, RC snubber circuits \cite{kellCompactFastMagnetic2021}, or
 high-voltage supplies\cite{uhthoff-rodriguezFastMagneticCoil2025b}.
In all cases, the rate of change of the current is limited by $d I/dt = V_\text{max}/L$, where $V_\text{max}$ depends on the details of the switching circuit.
Thus, for identical switching circuits, a Bitter-type coil with lower $L$ could have correspondingly quicker switching.



\section{\label{sec:thermal}Thermal Properties}


\subsection{\label{ssec:water} Water Cooling}

\begin{figure}
    \centering
    \includegraphics[width=\linewidth]{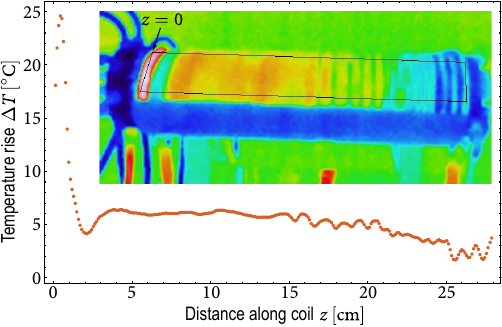}
    \caption{Longitudinal temperature profile of the ZS coil $t=36\,$s after switching on a DC current of 200 A. 
    Chilled water flows at 3~l/min with $T=20~^\circ$C---the coil temperature rise is shown in orange points relative to this temperature. 
    Temperatures are measured using a thermal imaging camera (image in inset), with a trapezoidal region defined for the coil.
    The lower blue region in the image is a plastic holder for the coil, and the cooling water lines are visible on the left and right ends of the image.
    Pixel values in this region are averaged in the transverse direction to obtain the orange points. 
    The temperature rise is roughly constant over most of the coil, with slightly lower temperature near the thicker, larger-spacing end.
    The hottest part of the coil is a localized region near $z=0$, corresponding to the orange data points in Fig.~\ref{fig:Tovertime}, which is likely due to a single faulty layer-spacer contact.}
    \label{fig:Tprofile}
\end{figure}

As seen in Fig.~\ref{fig:bfield-profile}, the ZS achieves the required field profile with a current of $200$~A, which corresponds to an ohmic power dissipation of $P=I^2R\sim1$~kW.
Compared to wire-wound coils of similar size\cite{mitraExploringAttractivelyInteracting2018}, this is about a factor of 5 increase in power.
Thus, water-cooling the coil is necessary in order to avoid excessive thermal stresses that could compromise its performance or risk damage to nearby components. 

We measured the temperature of the coil with cooling water at $T=20^\circ$\,C and a flow rate of 3~l/min using an infrared camera\footnote{Teledyne FLIR E6 Pro}. After running $200$~A of current through the ZS continuously for 36~s (2-3 times longer than the typical operation time in experiments), we discovered that the coil had a non-uniform temperature profile (see Fig.~\ref{fig:Tprofile}).
Near the low-field end of the coil, we saw that the temperature had increased the least, due to the lower resistance of the individual layers, as well as the larger inter-layer volume for cooling water.
We also saw a few layers with significantly higher temperature---these hot layers were likely caused by a single faulty layer--spacer contact, which dominated the total resistance of the coil and led to heating in the neighboring layers.

\begin{figure}[ht]
    \centering
    \includegraphics[width=\linewidth]{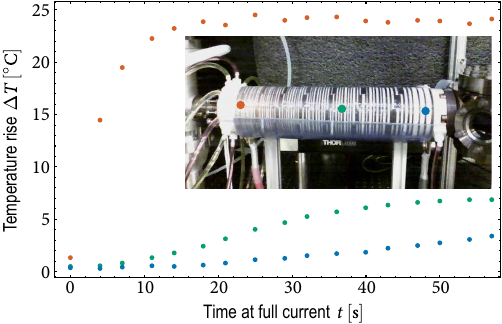}
    \caption{Temperature rise $\Delta T$ above the chilled water temperature (20~$^\circ$C at 3~l/min) as a function of time $t$ at 3 representative spots on the ZS coil during operation at $I=200$~A (shown on a photo of the coil in the inset). Temperature values are pixel-averaged transverse to the coil axis. The red dots measure the hottest region of the ZS coil, with $\Delta T\rightarrow 25~^\circ$C, reaching steady state the fastest, with a time constant $\tau \approx 5$~s. For the middle of the coil, shown in green dots, $\Delta T\rightarrow 7~^\circ$C with $\tau \approx 30$~s. For the coolest part of the coil, shown in blue dots, $\Delta T\rightarrow 5~^\circ$C with $\tau > 30$~s.}
    \label{fig:Tovertime}
\end{figure}

We measured the time-dependence of the temperature at three locations in the coil by taking an image with an infrared camera every few seconds over a minute, with an exposure time of $1/39$~s per image. The results are shown in Fig.~\ref{fig:Tovertime}.
We observed that the different parts of the coil reached different steady-state temperatures and on different timescales---about 10~s for the hottest part of the coil,  about 50~s for the middle of the coil, and longer than a minute for the coolest part of the coil.

Taking a temperature rise of $\Delta T=5^\circ$\,C as a rough average over the coil, we compute a thermal resistance $R_T=\Delta T/P=5^\circ$\,C$/$kW, similar to that of other similar designs\cite{sabulskyEfficientContinuousdutyBittertype2013}.

\section{\label{sec:conclusion}Conclusion}

We have described the design, construction, and characterization of a Bitter-type electromagnet with water cooling for the production of a magnetic field profile suitable for Zeeman slowing of atomic lithium. Our configuration has properties that are favorable for laser-cooling experiments---a lower resistance and self-inductance than traditional wire-wound designs---while requiring a single DC current supply and a modest amount of cooling water flow.
Nearly all of the components for its construction can be manufactured with a laser cutter and a 3-axis mill from commercial stock materials, and the coil can easily be baked along with a UHV chamber.

One drawback of this coil geometry is the large number of electrical contacts between the layers and spacers.
As such, a single faulty connection can dominate the coil resistance and lead to localized hot spots.
Another shortcoming of this design, and many traditional ZS designs, is that the coil is mechanically captured by the UHV chamber---in case of a cooling water leak or mechanical issue, the chamber vacuum would have to be broken to remove the coil for repair.

Future improvements to the design would simplify the construction of the coil, reducing the likelihood of individual bad contacts while also permitting easier repairs on the coil.
Such a modified ZS coil could simplify the setup for certain laser-cooling experiments, reducing the overall cost and hardware requirements.

\begin{acknowledgments}

We would like to thank Bart McGuyer for helpful discussions.
B.\,A.\,O. acknowledges support from the M.\,J.\, Murdock Charitable Trust, and the National Science Foundation through Grant No.\,PHY-2418777.

Initial measurements, calculations, and construction were carried out at Yale-NUS College of the National University of Singapore (NUS).

The data that support the findings of this study are openly available in Zenodo at \footnote{\href{https://doi.org/10.5281/zenodo.18779667}{https://doi.org/10.5281/zenodo.18779667}}.
\end{acknowledgments}










\bibliography{2025_BZS}

\end{document}